\documentclass[aps,prl,twocolumn,superscriptaddress]{revtex4}
\usepackage{graphicx} 
\usepackage{dcolumn}
\usepackage{bm}

\begin{document}

\title{Memory Effect in the Photoinduced Femtosecond Rotation
           of Magnetization \\ in the Ferromagnetic Semiconductor
           GaMnAs}

\author{J.~Wang}
\affiliation{Materials Sciences Division,
E.O. Lawrence Berkeley National Laboratory and Department of Physics, 
University of California
at Berkeley, Berkeley, California
94720, U.S.A.}

\author{I. Cotoros}
\affiliation{Materials Sciences Division,
E.O. Lawrence Berkeley National Laboratory and Department of Physics, 
University of California
at Berkeley, Berkeley, California
94720, U.S.A.}

\author{X. Liu}
\affiliation{Department of Physics, University of Notre Dame, Notre Dame, Indiana 46556, U.S.A.}

\author{J. Chovan}
\affiliation{Institute of Electronic Structure and Laser, Foundation for Research
and Technology-Hellas and Department of Physics, University of Crete, Heraklion, Greece.}

\author{J. K. Furdyna}
\affiliation{Department of Physics, University of Notre Dame, Notre Dame, Indiana 46556, U.S.A.}

\author{I. E. Perakis}
\affiliation{Institute of Electronic Structure and Laser, Foundation for Research
and Technology-Hellas and Department of Physics, University of Crete, Heraklion, Greece.}

\author{D. S. Chemla}
\affiliation{Materials Sciences Division,
E.O. Lawrence Berkeley National Laboratory and Department of Physics, 
University of California
at Berkeley, Berkeley, California
94720, U.S.A.}

\date{\today}

\begin{abstract}

We report a femtosecond response in photoinduced magnetization rotation 
in the ferromagnetic semiconductor GaMnAs, which allows for detection 
of a four-state magnetic memory at the femtosecond time scale. 
The temporal profile of this cooperative magnetization rotation 
exhibits a discontinuity that reveals
two distinct temporal regimes, marked by the transition from a highly
non-equilibrium, carrier-mediated regime within the first 200 fs, to a thermal, lattice-heating
picosecond regime.

\end{abstract}
\maketitle


\begin{figure}[floatfix]
\begin{center}
\includegraphics [scale=0.39] {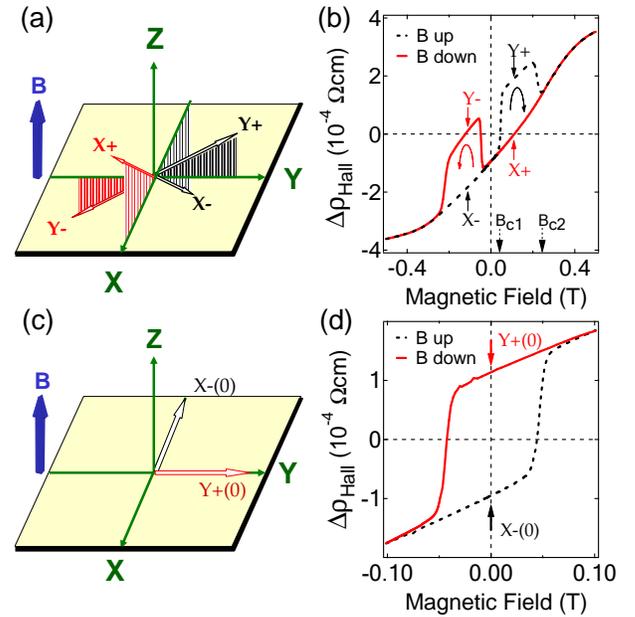}  
\caption{(Color online) Static magnetic memory. 
(a)-(b): Sweeping a slightly tilted {\em B} field (5$^o$ from 
the {\em Z}-axis and 33$^o$ from the {\em X}-axis) 
up (dashed line) and down (solid line) 
leads to 
consecutive 90$^o$ magnetization switchings 
between the {\em XZ} and {\em YZ} planes, 
manifesting as a ``major" hysteresis loop in the Hall magneto-resistivity. 
(c)-(d): ``Minor" hysteresis loop with {\em B} field sweeping in the
vicinity of 0T.  The magnetic memory state X$-$(0) or Y$+$(0) is 
parallel to one 
of the easy axis directions in the {\em XY} plane.
}
\end{center}
\label{dep}
\end{figure}

Magnetic materials displaying {\em carrier-mediated} ferromagnetic order offer
fascinating opportunities for non-thermal, potentially {\em femtosecond}
manipulation of magnetism.
A model system of such materials is Mn-doped III-V ferromagnetic 
semiconductors that have received a lot of attention lately \cite{ohno1998}.
On the one hand, their magnetic properties display a strong response 
to 
excitation with light or electrical gate and current
via carrier density tuning 
\cite{Koshiharaetal97PRL,ohnoetal2000,wangetalPRL2007}. 
On the other hand, the strong
coupling ($\sim$1 eV in GaMnAs) between carriers (holes) and Mn ions
inherent in carrier-mediated
ferromagnetism could enable a {\em femtosecond}
cooperative magnetic response induced by photoexcited carriers.
Indeed, the existence of a very early non-equilibrium,
non-thermal femtosecond regime of collective spin rotation in (III,Mn)Vs 
has been predicted theoretically \cite{ChovanetalPRL2006}. 
In addition, a coherent driving mechanism for femtosecond spin rotation 
via {\em virtual} excitations 
has also been recently 
demonstrated in antiferro- and ferri-magnets \cite{coherent}. 
Nevertheless, all 
prior studies of photoexcited magnetization rotation in 
ferromagnetic (III,Mn)Vs showed dynamics
on the few picosecond timescale, which accesses the quasi-equilibrium, quasi-thermal, 
lattice-heating regime \cite{psRotationGaMnAs}.
Up to now in these materials, 
the main observation on the femtosecond time scale 
has been photoinduced demagnetization 
\cite{wangetalPRL2005,wangetalreview2006,wang2008,Cywinski_PRB07}.
 
Custom-designed (III,Mn)V hetero- and nano-structures show rich 
magnetic memory
effects. One prominent example is GaMnAs-based four-state magnetic memory, 
where "giant" magneto-optical and magneto-transport
effects allow for ultrasensitive magnetic memory readout \cite{fourstate}.
However, all detection schemes demonstrated so far have been static 
measurements.
Achieving an understanding of collective magnetic phenomena 
on the femtosecond time scale is 
critical for terahertz detection of magnetic memory and therefore 
essential for developing realistic
"spintronic" devices and large-scale functional systems.

In this Letter, we report on photoinduced {\em femtosecond} collective 
magnetization rotation 
that allows for femtosecond detection of magnetic memory in GaMnAs. 
Our
time-resolved magneto-optical Kerr effect (MOKE) technique 
directly reveals a photoinduced 
four-state magnetic 
hysteresis via a quasi-instantaneous magnetization rotation.
We observe for the first time 
a distinct initial temporal regime of the magnetization rotation 
within the first $\sim$200 fs, 
during the photoexcitation
and highly non-equilibrium, non-thermal carrier redistribution times. We attribute the existence 
of such a regime to a {\em carrier-mediated} effective 
magnetic field pulse, arising without assistance from either
lattice heating or demagnetization.

The main sample studied was grown by low-temperature molecular beam epitaxy 
(MBE), and consisted of a 73-nm Ga$_{0.925}$Mn$_{0.075}$As layer on a 10 
nm GaAs 
buffer layer and a semi-insulating GaAs [100] substrate. The Curie temperature 
and hole density were 77 K and $3 \times 10^{20}$ cm$^{-3}$, respectively. 
As shown in Fig. 1,
our structure exhibits a four-state magnetic memory functionality.
By sweeping an external 
magnetic field B aligned nearly perpendicularly
to the 
sample normal, with small components in both the {\em X} and  {\em Y} 
directions in the sample plane, one can sequentially access four 
magnetic states, 
X$+\rightarrow$Y$-\rightarrow$X$-\rightarrow$Y$+$, via abrupt 90$^o$ 
magnetization ($\mathbf{M}$) switchings 
between the {\em XZ} and {\em YZ} planes [Fig. 1(a)]. 
In these magnetic 
states, $\mathbf{M}$ aligns along a direction arising as a combination of the 
external B field and the anisotropy fields, which point along the in-plane 
easy axes [100] and [010]. 
The multistep magnetic switchings manifest themselves as abrupt jumps in the 
four-state hysteresis in the Hall magneto-resistivity $\rho _{Hall}$
[Fig. 1(b)] (planar Hall effect 
\cite{fourstate}).
The continuous slopes of $\rho _{Hall}$ indicate a coherent out-of-plane $\mathbf{M}$ 
rotation 
during the perpendicular magnetization reversal (anomalous Hall effect 
\cite{ohno1998}).
Fig. 1(c)-(d) show the B scans in the vicinity of 0T, with the field turning 
points
 between the coercivity fields, i.e.,
$B_{c1}<\left|B\right|<B_{c2}$. This leads to a "minor" hysteresis loop, 
accessesing 
two magnetic memory states at $B=$0T: X$-$(0) and Y$+$(0).

We now turn to the transient magnetic phenonmena. 
We performed time-resolved MOKE spectroscopy \cite{wangetalreview2006} 
using 100 fs laser pulses. The linearly polarized
($\sim$12 degree from the crystal 
axis [100])
UV pump beam was 
chosen at 
3.1 eV, 
with peak fluence $\sim$ 10$\mu$J/cm$^2$. A NIR beam 
at 1.55 eV, 
kept nearly
perpendicular to the sample  ($\sim$ 0.65 degree from the normal),  
 was used as  probe. 
The signal measured in this polar geometry reflects the out-of-plane 
magnetization component, M$_z$.

\begin{figure}
\begin{center}
\includegraphics [scale=0.48] {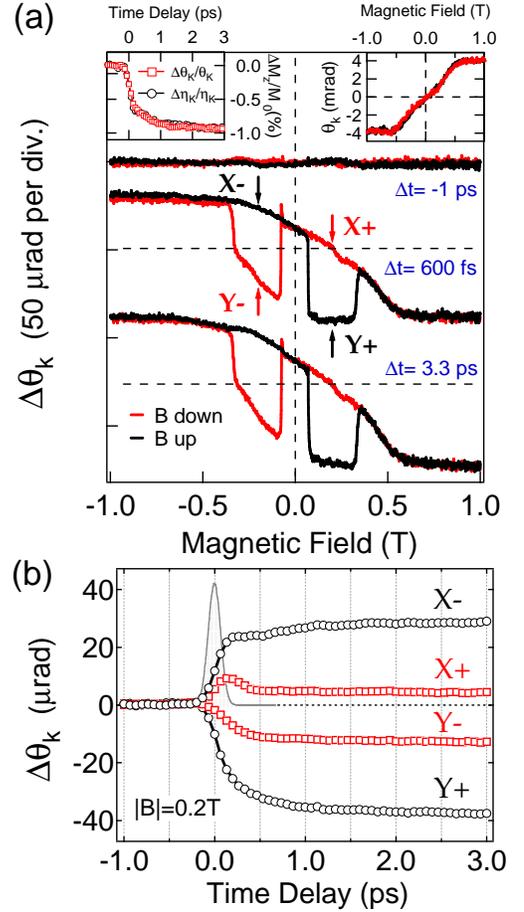}  
\caption{(Color) Photoinduced femtosecond four-state magnetic 
hysteresis.  
(a) B field scans of $\triangle \theta _{k}$ at 5K for
time delays $\triangle t=$ -1 ps, 600 fs, and 3.3 ps. 
The traces are vertically offset for clarity.  Inset (left): temporal profiles 
of normalized Kerr ($\theta _{k}$)  and 
ellipticity ($\eta _{k}$) angle changes 
at 1.0T; Inset (right): 
static magnetization curve at 5K ($\sim$4 mrad), measured in the same experimental 
condition (but without the pump pulse).
(b) Temporal profiles of photoinduced $\triangle \theta _{k}$ for the four
 magnetic states.  Shaded area:
pump--probe cross--correlation.}
\label{mag-dep}
\end{center}
\end{figure}
Fig. 2(a) shows the B field scan traces of the 
photoinduced change, $\Delta\theta_K$, 
in the Kerr rotation angle
at three time delays, $\triangle t=$ -1 ps, 600 fs, and 3.3 ps.
The magnetic origin of this femtosecond MOKE response 
\cite{KoopmansetAl00PRL} was confirmed by control
measurements showing a complete overlap of the pump--induced 
rotation ($\theta _{k}$)  
and ellipticity ($\eta _{k}$) changes [left inset, Fig. 2(a)].
$\Delta\theta_K$
is 
negligible at $\triangle t=$-1 ps.
However, a mere $\triangle t=$600 fs after photoexcitation, 
a clear photoinduced four-state 
magnetic hysteresis is observed in the magnetic field dependence 
of  $\Delta\theta_K$ (and therefore $\Delta M_z$),   
with four abrupt switchings 
at $\left|B_{c1}\right|=$0.074T and $\left|B_{c1}\right|=$0.33T due to the 
magnetic memory effects. 
As  
marked by the arrows in Fig. 2(a), the  
four magnetic states X$+$, X$-$, Y$-$, Y$+$ for 
$\left|B\right|=$0.2T give different
photoinduced  $\Delta\theta_K$.
It is critical to note that the steady-state 
MOKE curve, i.e. $\theta_K$  without pump field,
 doesn't show any sign of magnetic switching 
or memory behavior [right inset, Fig. 2(a)]; these 
arise from the pump photoexcitation. 
The B field scans also show a saturation 
behavior at $\left|B\right|>$0.6T,
to be discussed later.  We note that
the photo--induced hysteresis loops at $\triangle t=$3.3 ps
and 600 fs sustain similar shapes, 
with only slightly 
larger amplitudes at 3.3 ps. 
This observation confirms that the dynamic magnetic processes responsible 
for the abrupt switchings 
occur on a femtosecond time scale. 
Fig. 2(b) shows the photoinduced $\Delta\theta_K$ dynamics 
for the four initial states X$+$, X$-$, Y$-$, and Y$+$. An 
extremely fast $\Delta\theta_K$ develops within 200 fs, 
with magnitude and sign that distinctly differ, depending on the initially 
prepared state, consistent with Fig.2 (a). 
The substantial difference in  $\Delta\theta_K$
under the same B field 
- for instance between the X+ and Y+ states - 
shows that the magnetic dynamics is not due to simple demagnetization 
\cite{BeaurepaireetalPRL96, wangetalPRL2005, wangetalreview2006}.

The photoinduced dynamics of the zero--B field memory states [Fig. 1(c)] 
elucidates the salient features of the femtosecond magnetic processes. 
Fig. 3(a) shows the
temporal profiles of the photo-induced $\Delta\theta_K$ for 
X$-$(0) and Y$+$(0) initial states. 
Since the initial magnetization vector lies within 
the sample plane, $\Delta \theta _{k}$ in the first 200 fs 
reveals an out--of--plane spin rotation, with negligible contribution 
from demagnetization.  More intriguingly, the $\mathbf{M}$ in  X$-$
 and Y$+$ initial states rotates to different {\em Z}-axis directions, as
 illustrated in Fig. 3(b). This leads to opposite signs of the 
photoinduced signals and is responsible for the four-state
 magnetic switchings. 
Furthermore, the observation 
of an initial discontinuity in the 
temporal profiles of the $\mathbf{M}$ tilt reveals {\em two distinct
 temporal regimes}, marked in Fig. 3(a): a substantial magnetization
 rotation concludes after the first 200 fs and is followed by a {\em much 
slower} rotation change afterwards (over 100's of ps).

\begin{figure}
\begin{center}
\includegraphics [scale=0.4] {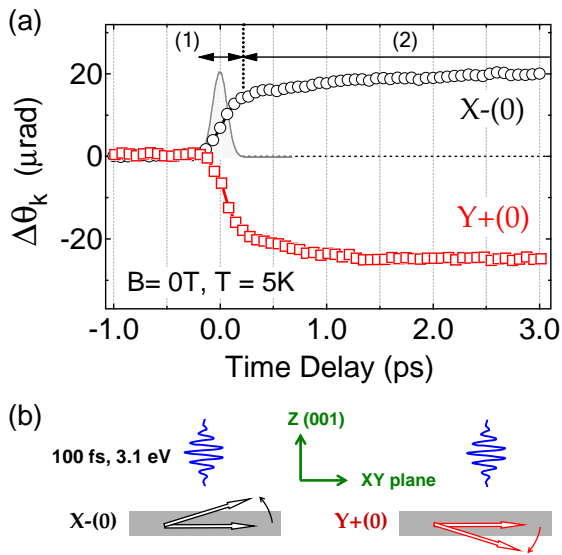}  
\caption{(Color online) (a) 
Photo-induced $\Delta \theta _{k}$ for 
two in-plane magnetic memory states, 
shown together with the pump-probe 
cross-correlation (shaded). The opposite, 
out--of--plane $\mathbf{M}$ rotations 
for the X$-$(0) and Y$+$(0) are illustrated in (b).}
\label{power}
\end{center}
\end{figure}

We now discuss the origin of the observed
femtosecond magnetization rotation. In the previously 
held picture of light--induced magnetization rotation in ferromagnets, 
the photoexcitation alters the anisotropy fields via 
quasi-equilibrium mechanisms, such as heating of the lattice 
(magneto-crystalline anisotropy) or heating of the spins (shape anisotropy) 
\cite{heating}. Since the in-plane magnetic memory states of Fig. 1(c) 
have negligible shape anisotropy, 
a significant 
B field within the standard picture 
can only occur on a time scale of several picoseconds
via the lattice heating mechanism. 
However, it has been shown theoretically 
\cite{ChovanetalPRL2006,ChovanetalPRB2008}
that the Mn spin in GaMnAs can
respond quasi-instantaneously to a femtosecond effective magnetic field 
pulse generated by hole spins via 
nonlinear optical processes assisted by interactions.
This light--induced B field pulse may be thought of as a femtosecond 
modification of the magnetic anisotropy fields.  
In the realistic system,
one needs to also treat microscopically the transient 
 magnetic anisotropy changes,
due to the complex valence bands and 
 highly non--thermal 
hole populations in the femtosecond regime,
 which 
drastically affect the photoexcited carrier spin.
Due to the {\em hole-mediated} effective exchange interaction between Mn 
spins, the 
anisotropy fields in GaMnAs result from the 
coupling of several valence bands by the {\em spin-orbit interaction} 
and depend on the transient hole distribution  
 and coherences between different bands \cite{ChovanetalPRB2008}.
In the static case, recent experimental \cite{anisotropy-exp} and 
theoretical \cite{anisotropy-the} investigations have shown that increasing 
the hole density significantly reduces the cubic anisotropy (K$_c$) along 
the [100] direction, while enhancing the uniaxial anisotropy (K$_u$) along
 [1-10]. One therefore expects that the photoexcited hole population 
turns on an effective magnetic field pulse
 ($\Delta B_{c}$) along the [1-10] direction [Fig. 4(a)].  This
photo-triggered $\Delta B_{c}$ then exerts a spin torque
 on the $\mathbf M$ vector, $\Delta \overrightarrow B_{c}\times 
\overrightarrow { M}$, and pulls it away from the sample plane.  The 
directions of these spin torques for the X$-$(0) and Y$+$(0) states are opposite, 
leading to different $\mathbf M$ rotation paths [Fig. 3]. 
Since this mechanism is mediated by the the non--thermal holes, 
the appearance of $\Delta B_{c}$ is 
quasi-instantaneous, limited only 
by the pulse duration of $\sim$100 fs \cite{ChovanetalPRB2008}. 
This femtosecond magnetic anisotropy
contribution from the non--thermal photoexcited carriers should be contrasted 
to the quasi--thermal contribution, arising from, e.g., the transient lattice
 temperature elevation on the picosecond time scale
\cite{psRotationGaMnAs}.  

Next we turn to the origin of the discontinuity that reveals the {\em two 
temporal regimes} in the collective magnetization rotation [Fig. 3]. The quick 
termination of the initial magnetization tilt implies that the
effective 
$\Delta B_{c}$ pulse induced by the photoexcitation decays within the 
first hundreds of femtoseconds.  The photoexcitation 
of a large (as compared to the 
ground state anisotropy field) $\Delta B_{c}$
requires an extensive {\em non-thermal} distribution of transient holes in {\em the 
high momentum states} of the valence band
\cite{anisotropy-the}. 
This is due to the large spin anisotropy of these hole states, 
empty in the unexcited sample,   
 via their
 strong spin-orbit interaction.
 In our experiment, immediately following 
 photoexcitation at 3.1 eV, a large density of transient holes distribute 
themselves over almost half of the Brillouin zone along the L[111] 
direction. The Mn--hole spin exchange interaction is also believed to
 be enhanced along [111] due to strong p-d orbital hybridization
 \cite{BurchPRB2004}. Consequently, 
these photoexcited holes contribute
 strongly to the magnetic anisotropy fields. The following rapid relaxation 
and thermalization of the high momentum holes,
 due to carrier-carrier and carrier-phonon scattering, reduce 
$\Delta B_{c}$ within a few hundred femtoseconds.
The subsequent picosecond magnetization rotation process 
arises from the change in magnetic anisotropy induced 
by the lattice temperature elevation. 
Our results reveal a complex scenario of collective spin
 rotation, 
marked by the transition from a non-equilibrium, carrier-mediated regime 
($<$200 fs) to a thermal, lattice-heating regime on the ps time scale.

\begin{figure}
\begin{center}
\includegraphics [scale=0.4] {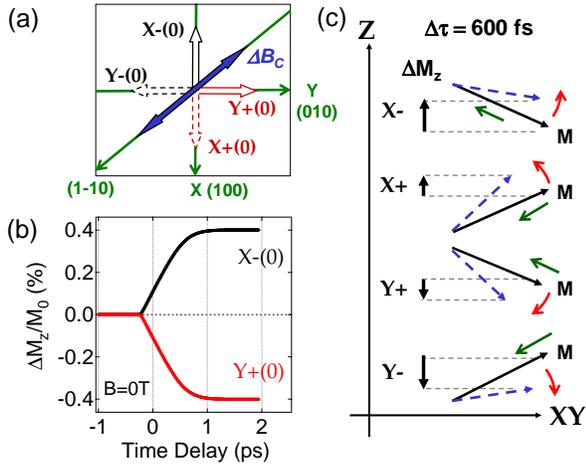}  
\caption{(Color online) (a) Schematics of the photoexcited carrier--induced 
anisotropy field $\Delta B_{c}$.  
(b) Simulations of $\Delta M_{z}/M_0$ for the two magnetic memory states.
 Parameters used in the calculation are $K_{c}=1.198\cdot 10^{-2}meV$, 
$K_{u}=0.373\cdot 10^{-2}meV$, $K_{3}=0.746\cdot 10^{-2}meV$, $T_{1}=330$fs 
and $3\% $ of
photoexcited carriers. (c) Schematics of the photoinduced M$_{z}$ for the X$-$,
 X$+$, Y$-$ and Y$+$ states at $\left|B\right|=$0.2T.
}
\label{temperature}
\end{center}
\end{figure}

We modelled the transient anisotropy 
 phenomenologically by 
deriving $\Delta B_{c}$ from the magnetic free energy, 
$$E_{anis}=-{K_{c}\over S^{4}}
S_{x}^{2}S_{y}^{2}+{K_{u}\over S^{2}}
S_{x}^{2}+{K_{3}\over S^{2}}
S_{z}^{2},$$ describing cubic (K$_c$) and uniaxial (K$_u$) 
contributions, and added a time--dependent modification of
$K_{c}/K_{u}$ due to the strongly anisotropic photoexcited hole states
\cite{ChovanetalPRB2008}. 
The corresponding contribution to the
Mn spin equation of motion is $\partial
_{t}{\bf S}={\bf S}\times {\bf H}_{anis}$, where ${\bf H}_{anis}=-
{\partial E_{anis}\over 
\partial {\bf S}}$. 
The light--induced change in the magnitude of $K_{c}/K_{u}$ 
increases during the pulse 
and then decreases with the energy relaxation time (T$_1$) 
of the high--momentum photoexcited holes. The results of our calculation 
are shown in Fig. 4(b), which gives a similar time dependence of the 
normalized $\Delta M_{z}$, with magnitude $\sim$ 0.4$\%$ of the 
total magnetization M$_0$ ($\sim$4 mrad at 5K), 
which compares well with the experiment. 


Finally, Fig. 4(c) illustrates the femtosecond detection of the 
four-state magnetic memory shown in Fig. 2.
By incorporating both the photo-induced rotation (red arrows) and the 
demagnetization (green arrows) effects, we can visualize the different 
M$_z$ changes for the four magnetic states, consistent with our observation. 
Demagnetization results in the high field saturation behaviour 
observed in Fig. 2(a).
For $\left|B\right|>0.60T$, $\mathbf M$ is aligned mostly along the sample 
normal. Then the photo-induced signals arise from the decrease in the 
$\mathbf M$ amplitude, 
which is more or less constant with respect to the field. 

In conclusion, we report on the femtosecond magnetic response of 
photoinduced 
magnetization rotation in GaMnAs, which allows for femtosecond detection of four-state 
magnetic memory.
Our observations unequivocally identify a {\em non-thermal, carrier-mediated} 
mechanism of magnetization rotation,  
relevant only in
the {\em femtosecond} regime, without assistance of
 either
lattice heating or demagnetization. This femtosecond cooperative magnetic
phenomenon may represent an as-yet-undiscovered universal principle in all
 carrier-mediated ferromagnetic
materials - a class of rapidly emerging ``multi-functional" materials with 
significant potential
for future applications, e.g., the oxides with promise of far above
room temperature Curie temperature.


This work was supported by the Office of Basic Energy Sciences of the 
US Department of
Energy under Contract No. DE-AC02-05CH11231, by the National Science
 Foundation DMR-0603752, and by the EU STREP program HYSWITCH.





\end{document}